\documentclass[%
onecolumn,
superscriptaddress,
amsmath,amssymb,
aps
]{revtex4-2}

\usepackage{wasysym}
\usepackage{hyperref}
\usepackage{lipsum}
\usepackage{csquotes}
\usepackage{graphicx}
\usepackage{dcolumn}
\usepackage{bm}

\usepackage{printlen} 
\usepackage{setspace}
\usepackage[capitalise]{cleveref}
\usepackage{physics}
\usepackage{xspace}
\usepackage[usenames,dvipsnames]{xcolor}
\usepackage[color=SeaGreen]{todonotes}

\newcommand{\vect}[1]{\mathbf{\boldsymbol{#1}}}
\newcommand{\tens}[1]{\mathbf{\underline{\boldsymbol{#1}}}}
\newcommand{\intd}{\,\mathrm{d}}

\usepackage{multirow}
\renewcommand\vec[1]{\vect{#1}}
\newcommand\mat[1]{\mathbf{#1}}
\newcommand{\st}[1]{^{\mathrm{#1}}}
\newcommand\tra{\st{T}}
\newcommand\sgn{\opbraces{\mathrm{sgn}}}
\newcommand\uvec[1]{\hat{\vec{#1}}}

\newcommand\cmpos{\ensuremath{\vec{r}\st{cm}}}
\newcommand\abij{\ensuremath{^{\alpha\smash{\beta}}_{ij}}} 
\newcommand\abii{\ensuremath{^{\alpha\smash{\beta}}_{ii}}}

\newcommand*{\kernhat}[2]{#2\kern#1\hat{\phantom{#2}}}

\let\oldchi\chi
\let\chi\undefined
\DeclareRobustCommand{\chi}{{\mathpalette\irchi\relax}}
\newcommand{\irchi}[2]{\raisebox{\depth}{$#1\oldchi$}} 

\newcommand{\cs}{~}

\newcounter{myfigpanel}[figure]
\newcounter{myfigpanelonly}[figure]

\newcommand{\panelletter}[1]{\refstepcounter{myfigpanel}\label{#1}\refstepcounter{myfigpanelonly}\label{onlyletter:#1}\alph{myfigpanel}}
\newcommand{\panel}[1]{(\protect\panelletter{#1})}

\crefalias{myfigpanel}{figure}
\crefname{myfigpanelonly}{panel}{panels}

\let\origcaption\caption
\let\caption\undefined
\DeclareRobustCommand{\caption}[1]{\origcaption{\protect\setcounter{myfigpanel}{0}\protect\setcounter{myfigpanelonly}{0}#1}}

\newcommand{\MPIDS}{\affiliation{Max Planck Institute for Dynamics and Self-Organization, Göttingen, Germany}}
\newcommand{\IDCS}{\affiliation{Institute for the Dynamics of Complex Systems, Göttingen University, Göttingen, Germany}}
\newcommand{\RPCTP}{\affiliation{Rudolf Peierls Centre for Theoretical Physics, University of Oxford, Oxford OX1 3PU, United Kingdom}}
\newcommand{\equalcontribution}{\thanks{these authors contributed equally}}

\begin{document}

\title{A minimal model of smoothly dividing disk-shaped cells}%

\author{Lukas Hupe}%
\equalcontribution
\MPIDS
\IDCS
\author{Yoav G. Pollack}%
\equalcontribution
\MPIDS
\IDCS
\author{Jonas Isensee}
\MPIDS
\IDCS
\author{Aboutaleb Amiri}
\thanks{current address: Carl Zeiss, Oberkochen, Germany}
\affiliation{Max Planck Institute for the Physics of Complex Systems, Dresden, Germany}
\author{Ramin Golestanian}
\email{ramin.golestanian@ds.mpg.de}
\MPIDS
\RPCTP
\author{Philip Bittihn}
\email{philip.bittihn@ds.mpg.de}
\MPIDS
\IDCS

\begin{abstract}
Replication through cell division is one of the most fundamental processes of life and a major driver of dynamics in systems ranging from bacterial colonies to embryogenesis, tissues and tumors. While regulation often plays a role in shaping self-organization, mounting evidence suggests that many biologically relevant behaviors exploit principles based on a limited number of physical ingredients, and particle-based models have become a popular platform to reconstitute and investigate these emergent dynamics. However, incorporating division into such models often leads to aberrant mechanical fluctuations that hamper physically meaningful analysis. Here, we present a minimal model focusing on mechanical consistency during division. Cells are comprised of two nodes, overlapping disks which separate from each other during cell division, resulting in transient dumbbell shapes. Internal degrees of freedom, cell-cell interactions and equations of motion are designed to ensure force continuity at all times, including through division, both for the dividing cell itself as well as interaction partners, while retaining the freedom to define arbitrary anisotropic mobilities. As a benchmark, we also translate an established model of proliferating spherocylinders with similar dynamics into our theoretical framework. Numerical simulations of both models demonstrate force continuity of the new disk cell model and quantify our improvements. We also investigate some basic collective behaviors related to alignment and orientational order and find consistency both between the models and with the literature. A reference implementation of the model is freely available as a package in the Julia programming language based on \textit{InPartS.jl}. Our model is ideally suited for the investigation of mechanical observables such as velocities and stresses, and is easily extensible with additional features.
\end{abstract}

\maketitle

Multicellular systems ranging from bacterial colonies to tissues, tumors, morphogenesis, and beyond exhibit a rich variety of self-organization phenomena and collective behavior.
These systems are driven out of equilibrium by different sources of activity, such as chemical reactions, signaling, metabolism, motility, and proliferation \cite{gompper_2020_2020, Hallatschek2023}. While many of these activities have mechanical consequences, proliferation – the growth and division of cells – is a particularly interesting case: Since it is an indispensable prerequisite for life, it must be relevant for any biological system at a certain stage of its life cycle. It involves, on a certain level of description, the creation and turnover of matter, often violating of number and volume conservation, thereby generating internal stresses and large-scale flows.
Proliferation usually enters the mathematical description at a fundamental level, making the violation of common assumptions a challenge. Besides the direct relevance of mechanical stresses for regulating proliferation\cs\cite{Moruzzi2021,Dolega2021,Minc2009,Kaukonen2016,Delarue2014,Montel2012}, many examples of self-organization in multicellular systems arise from mechanical interactions, such as orientational order\cs\cite{isensee_stress_2022,volfson_biomechanical_2008,dellarciprete_growing_2018,you_geometry_2018,langeslay_microdomains_2023,you_confinementinduced_2021}, topological defects\cs\cite{doostmohammadi_defectmediated_2016,Endresen2021,Guillamat2022,Ienaga2023,Lng2024}, collective motion\cs\cite{Beaune2018,Raghuraman2022,Luo2023,Lng2024}, to only name a few recent examples.
Given the plethora of interesting phenomena that can already be observed in relatively abstract physical models, a minimal yet consistent mathematical description of the mechanics during growth and division is therefore important to characterize these behaviors.
We therefore focus here on the theoretical and numerical underpinnings of such a mechanically consistent model instead of investigating a particular phenomenon, although we simulate some paradigmatic examples of collective behavior.

Depending on the level of detail and the type of questions addressed, multicellular systems can be modeled in different ways in order to study the collective dynamics.
Theoretical models often use coarse-grained descriptions such as continuum fields which allow analytical calculations and approximations of some observables such as order parameters, or scaling laws, which aid the construction of phase diagrams and studies of phase behavior \cite{gelimson_RG_2015,mahdisoltani_coarse-graining-field-theory_2021,BenAliZinati_2022,Wang_biofilm_2017}. Examples are continuum models of growing active nematics\cs\cite{dellarciprete_growing_2018,doostmohammadi_defectmediated_2016} and scaling theories for tissue growth and regeneration\cs\cite{Werner2015,gelimson_RG_2015,Ambrosi2019,Capek2019}.
Vertex models represent the system as a network of vertices connected by edges and implicitly assume confluency, where the dynamics of cell area, perimeter, and junctions are usually determined by an effective free energy, plus possible active contributions\cs\cite{Alt2017}.
They have been employed to elucidate, for example, the dynamics of in epithelial tissues\cs\cite{Barton2017}, tumor invasion\cs\cite{Li2019} and jamming transitions\cs\cite{Bi2016}.
Another type of abstraction is the division of available space into a lattice, leading to cellular-automaton-type models such as the cellular Potts model, to only name one popular example, which has been employed to model tumor growth\cs\cite{Szab2013} and morphogenesis\cs\cite{Hirashima2017}.
In particle-based models, each particle corresponds to a cell or a part of a cell, and interacts with other particles through explicit forces or potentials. A spectrum of models of different complexity and spirit have been used to investigate multicellular systems.
Recent directions include the extension of active-brownian-particle-like models with attraction and repulsion to model cohesive cell monolayers\cs\cite{Sarkar2021}, with contact inhibition of locomotion leading to emergent structures\cs\cite{Smeets2016}, entirely athermal models incorporating non-trivial internal cell cycle dynamics\cs\cite{Li2021} or nearly incompressible cells that are able to recapitulate nematic properties and large-scale orientational order in bacterial colonies\cs\cite{you_geometry_2018,isensee_stress_2022}. These kinds of extensions are easily possible in particle-based models, because they rely less on implicit quantities and effective energies than, e.g., vertex models or continuum models.
Independently of any additional mechanisms, however, these models always require a ``mechanical backbone'', which defines how different cells interact sterically. This backbone is what we focus on in this study for the case of growth and division, which inevitably have mechanical consequences.

In principle, because of their explicit formulation, incorporating explicit growth and division processes in particle-based models is straight-forward (as shown in some of the examples above).
However, these processes also come with unique challenges.  Most importantly, traditional methods of implementing the process of division often involve instantaneous and abrupt changes in the system, such as the sudden insertion of particles in new locations\cs\cite{ranft_fluidization_2010,aland_vertexparticle_2015,malmi_glasstumor_2018}, resulting in a discontinuity of forces, or the loss of cell identity.
These changes can affect the measurement of mechanical observables, such as stresses, strains, or pressures, and the tracking of cells across divisions, which are important for understanding the collective behavior and self-organization of the system in the context of non-equilibrium statistical physics.
Here, we build a particle-based model of dividing disk-shaped cells with an emphasis on the smoothness of all involved processes. Our goal is to avoid these unphysical mechanical fluctuations as much as possible.

As in many particle-based models of proliferation, we assume that cells always start from the same shape immediately after birth, which is chosen here to be radially symmetric for simplicity (corresponding to an aspect ratio of 1).
This inevitably requires some degree of transient anisotropy, which may appear already during the growth phase or, at the latest, at division.
In the first case, the anisotropy manifests as the direction of elongation and thus an anistropic shape change of the cell towards aspect ratio 2, before it divides into two particles of the original shape.
In the second case, all anisotropy is contained in the choice of the division plane, i.e. the axis along which the two daughter cells are placed. However, in this second case, new particles must necessarily be inserted in new places, causing force discontinuities which we are trying to avoid.
Hence, we investigate the first case of gradually introducing cell shape anisotropy.

First, we introduce a common set of coordinates and equations of motion which can be used to represent various physical interaction rules for objects with nematic symmetry.
It is based on two \emph{nodes} at opposite poles of the cell around which specific force laws can be built. Using this formulation and an appropriate force decomposition, we naturally ensure consistent motion of the entire cell as a rigid body in response to external forces and torques, while having the freedom to define arbitrary internal dynamics of the separation between nodes.
We then define two sets of interaction forces between the two nodes of a cell and between nodes of different cells: One is our new model of \emph{disk cells}, which treats individual nodes as disk-shaped elastic objects, and, as a consequence, the entire cell as a dumbbell during elongation.
With a few custom modifications, it is possible to achieve a high degree of force continuity in this case. As a comparison, the other set of interaction forces represents cells as spherocylinders---a common choice for rod-shaped bacteria in the literature---which also elongate incrementally and simplify to disks for aspect ratio 1.

In the second part of the study, we examine and compare both models: We start on the microscopic scale by demonstrating the presence or absence of force continuity and characterize force fluctuations in small growing colonies. We then extend the comparison to collective properties that manifest from these microscopic interactions, using cell orientation and nematic order as a prominent example.
Finally, we conclude by discussing how the unique features of our model will facilitate certain applications and by pointing out possible extensions.

\section{Disk cell model}
\label{sec:disks}

As the simplest description of an anisotropic particle, we model an individual cell $i$ as two \emph{nodes} connected by a spring of length $b_i$ and orientation $\varphi_i$, with a center of mass position $\cmpos_i$. \Cref{pan:model_rod_labelled} shows these degrees of freedom together with the two sets of shapes---disks (colored) and spherocylinders (dashed)---we will later introduce through appropriate interaction functions.

\subsection{Growth and division}
\begin{figure}[t]
\centering
\includegraphics{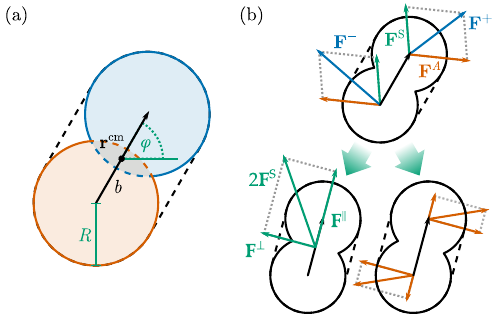}
\caption{%
\panel{pan:model_rod_labelled}~Illustration of a single rod-shaped cell, with backbone of length $b$, orientation $\varphi$, centre of mass position $\vec{r}\st{cm}$ and radius $R$.
\panel{pan:model_forcedecomposition}~Illustration of force decomposition. The symmetric (green, upper panel) and antisymmetric (orange, upper panel) components of the node forces (blue, upper panel) are decomposed into the centre-of-mass forces (green, lower panel) and the rotational and internal forces (orange, lower panel).
}
\end{figure}

To model the cellular life cycle, every cell has an internal clock $g_i \in [0, 1)$ which grows linearly in time with a growth rate $\dot g_i=\gamma_i$.
Growth is modeled by changing the rest length $b\st{eq}$ of the internal spring as a function of the internal clock, with
\begin{equation}\label{eq:model_disk_restlength} b\st{eq}(g_i, R) = 2R \cdot g_i\,.
\end{equation}
At $g_i = 0$, both nodes overlap completely, while at $g_i = 1$ they tend to move completely apart.
Cells can now be divided into two equally sized daughter cells, each initialised with internal clocks set to zero, in the positions of the two nodes of the parent cell.
To desynchronize growth cycles, we draw the growth rates of these newly born cells independently from a growth rate distribution $P(\gamma)$.
In the following, we will choose growth rates from a uniform distribution on the interval $[0.75\gamma_0, 1.25\gamma_0]$ with a constant $\gamma_0$.
The cell orientation as well as all other static parameters are inherited from the parent.

An overview of all static and dynamic model parameters as well as some of the functions and symbols used to describe the model can be found in Table~\ref{tab:model_rod_overview}.

\subsection{Equations of motion}

We model the dynamics of the cells with overdamped equations of motion, i.e.
\begin{equation}
\label{eq:model_overdamped}
\dv{t} \vec{x} = \tens{\mu}\cdot\sum\limits_\alpha\vec{F}_\alpha \;,
\end{equation}
where $\mat{\mu}$ is the mobility tensor.

For a single cell with the four degrees of freedom $\vec{r}\st{cm}_i$, $b_i$ and $\varphi_i$, we can write these equations as
\begin{align}
\label{eq:model_generic_eom_cmpos}
\dv{t}\,\vec{r}\st{cm}_i &= \tens{\mu}\st{cm}(b_i, \varphi_i) \cdot \vec{F}\st{cm}_i\\
\label{eq:model_generic_eom_int}
\dv{t}\,b_i &= \mu\st{int}(b_i)\cdot F\st{int}_i\\
\label{eq:model_generic_eom_rot}
\dv{t}\,\varphi_i &= \mu\st{rot}(b_i)\cdot T_i.
\end{align}
where $\vec{F}\st{cm}_i$ is the force acting on its center of mass, $F\st{int}_i$ is the force acting on the internal spring and $T_i$ is the torque with respect to the center of mass.
We can simplify Eq.~\eqref{eq:model_generic_eom_cmpos} by decomposing the right-hand side into its components parallel and perpendicular to the backbone.
This effectively diagonalizes $\mat{\mu}_{cm}$, eliminates the angle dependency and yields
\begin{equation}
\label{eq:model_generic_eom_cmpos_decomposed}
\dv{t}\,\vec{r}\st{cm}_i = \mu^{\parallel}(b_i)\;F^{\parallel}_i\cdot\uvec{e}_i + \mu^{\perp}(b_i)\;F^{\perp}_i\cdot\qty(\uvec{z}\times\uvec{e}_i)\\
\end{equation}
where $\uvec{e}_i = (\cos(\varphi_i), \sin(\varphi_i))\tra$ and the two-dimensional ``cross product'' is defined as $\uvec{z}\times\qty(x_1, x_2)\tra = (-x_2, x_1)\tra$.

Our implementation of the model allows for different choice of mobilities, enabling us to model different sources of friction in different applications.
Here, we use a numerical approximation for the mobilities of a dumbbell in a viscous fluid, derived by \citeauthor{luders_microscopic_2021}~\cite{luders_microscopic_2021}, to set the external mobilities $\mu^{\parallel}$, $\mu^{\perp}$ and $\mu\st{rot}$ as
\begin{alignat}{5}
\label{eq:model_mob_par}
\mu^\parallel(a) &= \frac{1}{2\pi\,\eta\,\qty(2R\,a)}  \,\Big(&\log(a) &-0.0552 &+&\,0.8477&\,a^{-1}&-&0.1254 &\,a^{-2}\Big)\\
\label{eq:model_mob_perp}
\mu^\perp(a)     &= \frac{1}{4\pi\,\eta\,\qty(2R\,a)}  \,\Big(&\log(a) &+1.025  &+&\,0.1317&\,a^{-1}&+&0.178  &\,a^{-2}\Big)\\
\label{eq:model_mob_rot}
\mu\st{rot}(a)   &= \frac{3}{ \pi\,\eta\,\qty(2R\,a)^3}\,\Big(&\log(a) &-0.3429\,&+&\,0.7749&\,a^{-1}&-&\,0.09898&\,a^{-2}\Big)
\end{alignat}
where $a = b_i/2R + 1$ is the aspect ratio of the cell.

To choose a value for the internal mobility $\mu\st{int}$ that is consistent with the external mobilities, we first need to consider how external forces can act on the internal degree of freedom.

\subsection{Force decomposition}
\label{sec:force_decomposition}
An external force acting on a cell can---depending on its impact point and angle---influence all four degrees of freedom (position, orientation, backbone length) simultaneously.
For the disk model, all steric forces will be modeled as central forces acting on the two nodes located at the positions
\begin{equation}
\label{eq:model_rod_nodes}
\vec{r}^\pm_i = \cmpos_i \pm \frac{b_i\,\uvec{e}_i}{2}.
\end{equation}
We call $\vec{r}^\pm_i$ the \emph{positive} and \emph{negative node} of the cell.
Later, we will define all interactions in terms of the force $\vec F\abij$ exerted on the node $\alpha \in \{+, -\}$ of a cell $i$ by the node $\beta$ of another cell $j$, which we assume to be a function of the inter-node distance vectors.
We define the corresponding distance vector from node $\beta$ of cell $j$ to node $\alpha$ of cell $i$ as $\vec d\abij = \vec r^\alpha_i - \vec r^{\smash{\beta}\vphantom{\alpha}}_j$.
This choice means that the backbone vector connecting two nodes of the \emph{same} cell can be expressed as $b_i\uvec e_i=\smash{\vec{d}^{+ -}_{ii}}$. It also means that repulsive forces $\smash{\vec F\abij}$ are always parallel to their respective distance vectors $\smash{\vec d\abij}$.
However, as the equations of motion require forces in cell coordinates, at some point the total force $\smash{\vec{F}^\alpha_i = \sum_{\beta, j}\vec{F}\abij}$ exerted on the node $\alpha$ of cell $i$ by all other nodes has to be transformed back into cell coordinates.
For this, the first step is to decompose the forces into a symmetric and an antisymmetric component (compare Figure~\ref{pan:model_forcedecomposition}.)
\begin{align}
\label{eq:model_force_sym}
\vec{F}\st{S}_i &= \frac{1}{2}\qty(\vec{F}^+_i + \vec{F}^-_i)\\
\label{eq:model_force_antisym}
\vec{F}\st{A}_i &= \frac{1}{2}\qty(\vec{F}^+_i - \vec{F}^-_i).
\end{align}

The force experienced by the center of mass must be equal to the sum of the forces experienced by the nodes, i.e.\@ $2\vec{F}\st{S}_i$.
Using the scalar components $F^\parallel_i$ and $F^\perp_i$ of the center of  mass force (compare Eq.~\eqref{eq:model_generic_eom_cmpos_decomposed}), we can write
\begin{align}
\label{eq:model_rod_fpar}
F^\parallel_i &= \uvec{e}_i\cdot 2\,\vec{F}\st{S}_i\\
\label{eq:model_rod_fperp}
F^\perp_i &= \qty(\uvec{z}\times\uvec{e}_i)\cdot{}2\,\vec{F}\st{S}_i
\end{align}
with the cross product $\uvec{z}\times\uvec{e}_i$ as defined in Eq.~\eqref{eq:model_generic_eom_cmpos_decomposed}.

An analogous decomposition can be applied to the antisymmetric component:
The parallel antisymmetric force will affect the cell length, while the orthogonal antisymmetric force will change the cell orientation.
This can be expressed as a scalar force and a torque
\begin{align}
\label{eq:model_rod_fint}
F\st{int}_i &= \uvec{e}_i\cdot 2\,\vec{F}\st{A}_i\\
\label{eq:model_rod_torque}
T_i &= \uvec{z}\cdot\left(b_i\uvec{e}_i\times\vec{F}\st{A}_i\right).
\end{align}
Note that the factor of two in Eq.~\eqref{eq:model_rod_fint} is a matter of convention and can be compensated for when choosing the internal mobility.

We can now examine the effect of the value of the internal mobility $\mu\st{int}$ on the dynamics of the node positions $\vec{r}^\pm$ by considering a simple example:
A force $\vec{F}^-_i = f\,\uvec{e}_i$ is applied to the negative node $\vec{r}^-_i$ of a cell at rest ($b_i = b\st{eq}(g_i)$).
In the absence of explicit coupling between the two nodes (i.e.\@ without the internal spring), the positive node remains force-free.
Using the force decomposition as outlined above, we find $F^\parallel_i = f$ and $F\st{int} = -f$.
We can thus write the velocity of the positive node as
\begin{align}
\nonumber
\dot{\vec{r}}^+_i &= \dot{\vec{r}}\st{cm}_i + \frac{\dot{b_i}\,\vec{e}_i}{2}\\
\label{eq:model_rod_asym_push}
&= \frac{2\mu^\parallel - \mu\st{int}}{2}\,f\,\vec{e}_i\;.
\end{align}
From this result, we see that for $\mu\st{int}\ne2\mu^\parallel$, there is an implicit coupling between the nodes that will cause the positive node to move without any force acting on it.
In fact, if $\mu\st{int} > 2\mu^\parallel$, it will move in the opposite direction to the force applied to the cell.
Thus, we only consider $\mu\st{int} \le 2\mu^\parallel$ physically meaningful and choose $\mu\st{int} = 2\mu^\parallel$ as the canonical value.

\subsection{Interactions}
In real life the interactions between cells in a colony or tissue can be fairly complex and include diverse effects such as adhesion and chemical signalling.
Although these could be implemented in principle, we limit ourselves here to mechanical repulsion caused by volume exclusion, which we model using pairwise forces between the nodes: Two overlapping nodes repel each other with  force from Hertzian contact theory for homogeneous elastic spheres, which scales with the $\flatfrac{3}{2}$th power of the overlap~\cite{hertz_uber_1882, willert_handbook_2019}.
We can write the force exerted on the node $\alpha$ of a cell $i$ by the node $\beta$ of another cell $j$ as
\begin{equation}
\label{eq:model_disk_force}
\vec{F}\abij = m_{ij}\frac{Y}{2}\sqrt{\frac{R}{2}}\qty(2R - \norm{\vec{d}\abij})^{3/2}  \frac{\vec{d}\abij}{\norm{\vec{d}\abij}}\text{ for }i\neq{}j,\text{ and }\norm{\vec{d}\abij} \le 2R
\end{equation}
Here, $m_{ij}$ is a \emph{softness factor}, which serves to compensate for instantaneous node doubling on division (see further below).
The additional factors arise from the transformation of the normal contact problem of two elastic spheres to that of a single hard indenter in an elastic half-space:
For this, the hardness of the half-space is computed from the effective Young's moduli\cs\footnote{Technically, these also depend on the Poisson's ratios of the two materials. Since we do not fully model the elastic properties of the cells, we can neglect this intricacy here.}
(definition in~\cite[Eq. (2.1)]{willert_handbook_2019}) $Y_1$ and $Y_2$ of the two spheres, with
\begin{equation}
\frac{1}{Y^*} = \frac{1}{Y_1} + \frac{1}{Y_2}\,.
\end{equation}
For sphere radii $R_1$ and $R_2$, the curvature of the equivalent hard indenter at the contact point is the sum of the sphere curvatures, i.e.
\begin{equation}
\frac{1}{R^*} = \frac{1}{R_1} + \frac{1}{R_2}
\end{equation}
with an effective radius $R^*$.
Solving the Boussinesq problem for this geometry yields a normal force proportional to $E^*\,\sqrt{R^* \delta^3}$ for an indentation depth $\delta$~\cite[Section 2.5.3]{willert_handbook_2019}.
Assuming equal radii $R := R_1 = R_2$ and Young's moduli $Y := Y_1 = Y_2$, this simplifies to the expression in~\cref{eq:model_disk_force}.
The numerical implementation uses the full expression without this simplification, but here we restrict ourselves to the simpler case.

To keep steric forces between nodes continuous on division, we use the same Hertzian force law for the internal spring connecting the two nodes
\begin{equation}
\label{eq:model_disk_internalforce}
\vec{F}\abii = H\st{int}\frac{Y}{2}\sqrt{\frac{R}{2}}\cdot\sgn(\Delta b_i)\cdot\abs{\Delta b_i}^{3/2}\,\frac{\vec{d}\abii}{\norm{\vec{d}\abii}}\text{ for }\alpha\neq\beta
\end{equation}
with $\Delta b_i = b\st{eq}(g_i, R) - b_i$. $H\st{int}>0$ is an ``internal hardness factor'' which can be used to make the backbone length more or less flexible, but is always set to 1 in the disk cell model for the purpose of this study. Note that $\vec{d}^{+ -}_{ii}$ is simply the backbone vector $b_i\uvec e_i$ according to our index convention (see beginning of \cref{sec:force_decomposition}) and therefore $\norm{\vec d^{- +}_{ii}}=\norm{\vec d^{+ -}_{ii}}=b_i$ and $\vec d^{+ -}_{ii}/\norm{\vec d^{+ -}_{ii}}=-\vec d^{- +}_{ii}/\norm{\vec d^{- +}_{ii}}=\uvec e_i$ in \cref{eq:model_disk_internalforce}. Therefore, with $H\st{int}=1$, the functional form of the internal spring and the inter-cell force law in \cref{eq:model_disk_internalforce,eq:model_disk_force} are exactly matched at cell division, when $b\st{eq}(g_i=1, R) = 2R$. This leads to continuity in the forces when nodes switch from belonging to the same mother cell using the latter interaction rule to becoming two distinct daughter cells each with their own nodes using the former rule, independently of the cell's compression state. Note that this is not true for the case $b_i> b\st{eq}$ (i.e. the backbone is expanded instead of relaxed or compressed), since \cref{eq:model_disk_force} does not have an attractive regime, but this case rarely occurs and, if so, could be considered a physical rupture event for which the force jump is actually correct.
\begin{figure}[t]
\centering
\includegraphics{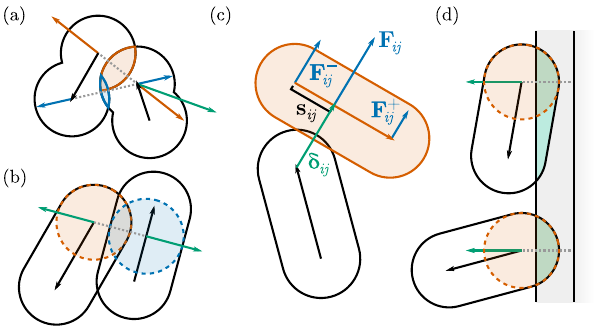}
\caption{%
\panel{pan:model_disk_interaction}~Two interacting disk cells with highlighted node overlaps and forces.
\panel{pan:model_rod_interaction}~Two interacting rod cells with highlighted virtual Hertzian disks and shortest distance $\delta_{ij}$ (grey dotted line).
\panel{pan:model_rod_forceattribution}~Illustration of the redistribution of a force $\vec{F}_{ij}$ onto the virtual node forces of cell $i$ (red).
\panel{pan:model_rod_overlap}~Two short rod cells interacting with much longer cell, showing that pivoting the cell around one of its nodes can cause the overlap area (shaded in green) to grow dramatically while keeping the forces constant.
}
\end{figure}

As cell division is implemented by spawning the child cells (containing two nodes in identical positions each) with their cell center positions at the nodes $\vec{r}^\pm$ of the parent cell, there is another potential source of discontinuity: While child cells take up the exact space occupied by the mother (due to their identical joint shape and the force law matching above), all forces due to overlaps with nodes of neighboring cells will double on division, as the single overlapping node is replaced by the two nodes of the corresponding child cell. Similarly, the forces between the two nodes of the same cell would quadruple when they are replaced by two nodes each in identical positions upon division.
To prevent these discontinuities, we include a softness factor $m_{ij}$ in force law in \cref{eq:model_disk_force} to continuously connect the two situations before and after division, and interpolate between them:
\begin{equation}
\label{eq:model_disk_softness}
m_{ij} = \frac{(g_i +1)\,(g_j + 1)}{4}.
\end{equation}
Each cell $i$ effectively contributes a factor $(g_i +1)/2$ to this expression, leading to half-strength forces for $g_i=0$ directly after division. For cells interacting with the two nodes of a child cell in identical positions, these then add up to the full-strength interaction force with the corresponding node of the mother $k$ that was present just before division (when $g_k=1$). In combination, for daughter cells $i$ and $j$ of the same mother cell, $m_{ij}$ exactly compensates the increase in the number of interactions from 1 internal one to 4 external ones. In between birth and division, $m_{ij}$ continuously interpolates between the two extremes, increasing the forces from half to full strength.

\begin{table}[t]
\centering\small\renewcommand{\arraystretch}{1.1}%
\begin{tabular}{ll}
\multicolumn{2}{c}{\scshape dynamic variables}\\\hline
$\vec{r}^\pm_i$&node positions\\
$\vec{r}\st{cm}_i$&centre of mass position\\
$b_i$&backbone length\\
$\varphi_i$&cell orientation\\
$g_i$&growth phase\\[1em]
\multicolumn{2}{c}{\scshape static parameters}\\\hline
$l\st{max}$&division length (fixed to $4R$ for disk cells)\\
$\gamma_i$&growth rate, $\gamma_i = \dv*{g_i}{t}$\\
$R$& node radius / cell extent from backbone\\
$Y$&Young's modulus\\
$H\st{int}$&internal spring hardness factor\\
$\eta$&viscosity\\
\end{tabular}\quad%
\begin{tabular}{ll}
\multicolumn{2}{c}{\scshape forces}\\\hline
$\vec{F}^\pm_i$&node forces\\
$F^\parallel_i$&\multirow{2}{*}{ translational forces}\\
$F^\perp_i$&\\
$F\st{int}_i$&internal force\\
$T_i$&torque\\[1em]
\multicolumn{2}{c}{\scshape functions}\\\hline
$b\st{eq}(g_i)$&backbone rest length\\
$\mu^\parallel(b_i)$& \multirow{2}{*}{translational mobilities}\\
$\mu^\perp(b_i)$&\\
$\mu\st{rot}(b_i)$&rotational mobility\\
$\mu\st{int}(b_i)$&internal mobility\\
$P(\gamma)$& growth rate distribution
\end{tabular}
\caption{Overview of model parameters. A subscript index \textit{i} indicates that the quantity is specific to an individual cell. In node-specific quantities, the superscript $+$ or $-$ indicates the values for the positive and negative node, respectively. The symbols $\parallel$ and $\perp$ denote vector components longitudinal or transverse to the cell.}
\label{tab:model_rod_overview}
\end{table}

\section{Rod cell model}
\label{sec:rods}
For comparison with a traditional model from the literature which also features incremental elongation, we now specify an alternative set of force laws and slightly modified equations of motion, using the same framework of two nodes (which we will call \emph{pseudo}nodes in the context of this model) and the corresponding force decomposition as defined for disk cells. The model (without the translation to our theoretical framework) has been commonly used in the literature for both two- and three-dimensional simulations of bacilliform bacteria~\cite{vanholthetotechten_defect_2020, storck_variable_2014, orozco-fuentes_order_2013, volfson_biomechanical_2008, cho_selforganization_2007, you_geometry_2018} and represents cells as discorectangles (i.e., the two-dimensional equivalent of spherocylinders), where the cell ``surface'' is defined as the locus of points with a distance of $R$ to the cell backbone, as indicated by the dashed outline in \cref{pan:model_rod_labelled}. Since it also exhibits gradual elongation with subsequent division and its parameter space includes the case of cells growing from aspect ratio 1 to 2 before dividing, it is comparable with our disk cell model in many aspects except the exact magnitude of the involved forces.

\subsection{Growth and division}
\label{sec:rods_growth_and_division}
Unlike disk cells, this model does not impose any upper limit on the division length $l\st{max}$, although we limit ourselves here to the case $l\st{max}=4R$ in analogy to the disk cells.
The time evolution of the rest length therefore depends on this parameter, with
\begin{equation}
\label{eq:model_rod_restlength}
b\st{eq}(g_i, l\st{max}, R) = \frac{l\st{max}}{2}\cdot (g_i + 1)  - 2R\,.
\end{equation}

As another difference to our disk cell model, the backbone length $b_i$ is commonly not considered a dynamic degree of freedom which can respond to external forces. In fact, our reference implementation provides the option to endow it with the same dynamics as for disk cells, according to \cref{eq:model_generic_eom_int,eq:model_disk_internalforce}. However, to stay consistent with the literature, in this study, we only consider the limit of an infinitely stiff internal spring $H\st{int}\rightarrow\infty$, such that the backbone length follows the equilibrium length
\begin{equation}
\label{eq:incompressible_backbonelength}
b_i\equiv b\st{eq}(g_i, l\st{max}, R).
\end{equation}
Enslaving the backbone length to the growth progress in this way corresponds to a rigid backbone which instantaneously balances any longitudinal antisymmetric forces which would usually compress it and change the cell length according to \cref{eq:model_rod_fint}. Without growth, this would imply a force of the backbone exerted on each pseudonode of
$\vec{F}\abii = -\frac{1}{2}(\uvec{e}_i\cdot\,\tilde{\vec{F}}\st{A}_i)\,\vec{d}\abii/\norm{\vec d\abii}\,$,
where $\tilde{\vec{F}}\st{A}_i$ is the antisymmetric force calculated without the contribution from the internal spring.
It can be easily checked that this force contribution leads to a vanishing longitudinal component of the antisymmetric force $\vec{F}\st{A}_i$ and therefore $F\st{int}_i=0$ according to \cref{eq:model_rod_fint}, consistent with a rigid backbone. For a growing cell, we have to add a contribution to the virtual force that causes the prescribed elongation according to \cref{eq:model_generic_eom_int}. Since we know that the resulting change rate is $\dot b_i = \frac{l\st{max}}{2}\dot g_i=\frac{\gamma_i \, l\st{max}}{2}$, we thus require $F\st{int}_i = \frac{\gamma_i \, l\st{max}}{2\mu\st{int}}$ (while the internal mobility $\mu\st{ing}$ can be chosen depending on our assumptions about the internal dynamics of the cell during elongation, we choose it here as $\mu\st{int} = 2\mu^\parallel$ for comparability with the disk cell model, see end of \cref{sec:force_decomposition}). It is easy to verify that this is achieved by assuming
\begin{equation}
\label{eq:incompressible_Fint}
\vec{F}\abii =  \left(-\frac{1}{2}(\uvec{e}_i\cdot\,\tilde{\vec{F}}\st{A}_i)  +  \frac{\gamma_i \, l\st{max}}{8\mu\st{int}}\right)\,\frac{\vec d\abii}{\norm{\vec d\abii}},
\end{equation}
which contains both contributions. It should be emphasized that calculating this force is not required for simulating the numerical model, since we simply prescribe the length of the backbone according to \cref{eq:incompressible_backbonelength} in the case of an incompressible backbone. However, it is important to take it into account when comparing node forces in simulations with $H\st{int}\rightarrow\infty$ to those with finite $H\st{int}$, as is the case below, when we will compare our disk cells ($H\st{int}=1$) to traditional rod cells ($H\st{int}\rightarrow\infty$).

For cell division, determining position and elongation of the children is slightly more complicated than in the disk model:
For arbitrary $l\st{max}$, we cannot always place new cells with their centres of mass at the parent's pseudonode positions.
Hence we have chosen to initialize child cells so that the sum of their total lengths (including the caps) is equal to the total length of the parent cell, without any additional overlap in between the cells. Note that, in the case of rod cells with $l\st{max}=4R$ and rigid backbones ($H\st{int}\rightarrow\infty$) as considered here, this strategy indeed simplifies to placing new cells with backbone length 0 (i.e., circular cells) at the parent cell's pseudo-node positions, just like for disk cells (one reason why a rod cell model serves as a good comparison here). However, both for the special case considered here and the more general case of arbitrary $l\st{max}$ and $H\st{int}$, this choice is merely a tradeoff, as there is no strategy which can preserve all internal and external forces across the division, as we will see explicitly in \cref{sec:forcecontinuity}.

\subsection{Interactions}
\label{sec:rods_interactions}
Forces between two cells $i$ and $j$ are modeled as functions of the shortest distance vector $\vec{\delta}_{ij}$ between their backbones as a first order approximation of their volume overlap integral, with the force $\vec{F}_{ij}$ acting on cell $i$ due to cell $j$ defined as
\begin{equation}
\label{eq:model_rod_external_steric}
\vec{F}_{ij}(\vec{\delta}_{ij}) = \frac{Y}{2}\sqrt\frac{R}{2}\,\qty(2R - \norm{\vec{\delta}_{ij}})^{3/2} \,\frac{\vec{\delta}_{ij}}{\norm{\vec{\delta}_{ij}}}
\end{equation}
for $\norm{\vec{\delta}_{ij}} < 2R$ and zero otherwise.
This effectively introduces two virtual Hertzian disks on the backbones at the points of closest approach, as illustrated in Figure~\ref{pan:model_forcedecomposition}.
Numerically, we find $\vec{\delta}_{ij}$ by using an algorithm\cs\cite{ericson_realtime_2004} that efficiently finds the relative positions $s_{ij}$ and $s_{ji}$ of the closest points along the backbones of cells $i$ and $j$, where we identify $s_{ij}=0$ with the position $\vec{r}^{-}_i$ of the negative pseudonode and $s_{ij}=1$ with the position $\vec{r}^{+}_i$ of the positive pseudonode according to \cref{eq:model_rod_nodes}, i.e., $s$ increases in the direction of $\uvec{e}_i$.

It should be noted that while this construction allows force to act at arbitrary positions along the backbone, they will always be orthogonal to it (except on the end points).
This allows us to efficiently transform the forces into comparable effective forces acting on the ends of the backbone (Figure~\ref{pan:model_rod_forceattribution}), thus enabling us to re-use the force decomposition methods built for the disk model~\footnote{Of course, torques and center-of-mass forces could also be calculated directly, but this allows us to treat disk and rod cells within the same framework and will facilitates quantitative comparisons below.}.
This is done by using the relative position $s_{ij} \in [0, 1]$ of the attack point on the backbone of cell $i$ (as returned by the closest points algorithm) and then distributing the forces onto the pseudonode components accordingly
\begin{equation}
\label{eq:model_rod_nodeforces}
\begin{aligned}
\vec{F}^{+}_{ij} &= s_{ij}\,\vec{F}_{ij}\\
\vec{F}^{-}_{ij} &= (1 - s_{ij})\,\vec{F}_{ij}.
\end{aligned}
\end{equation}

From this definition, we can already see that this transformation does not affect the centre of mass force of cell $i$, as
\begin{align}
\vec{F}\st{cm}_i &= \sum_j \vec{F}\st{+}_{ij} + \vec{F}\st{-}_{ij} = \sum_j \qty(s_{ij} + \qty(1 - s_{ij}))\vec{F}_{ij}\\
&= \sum_j \vec{F}_{ij}\;.
\end{align}

Preservation of torques is less obvious, but can be shown with a simple calculation:
A force $\vec{F}$ acting on the backbone at some fraction $s$ along its length will create a torque $\tau = \uvec{z}\cdot\left(\flatfrac{(2s-1)\,b_i\uvec{e}_i}{2} \times \vec F \right)$ with respect to the center of mass.
Distributing $\vec{F}$ onto the node forces and computing the torque $T$ using Eq.~\eqref{eq:model_rod_torque}, we get
\begin{align}
T &= \uvec{z}\cdot\qty(b_i\uvec{e}_i\times\qty(\frac{\vec{F}\st{+} - \vec{F}\st{-}}{2}))\nonumber\\
&= \uvec{z}\cdot\qty(b_i\uvec{e}_i\times\qty(\frac{2s - 1}{2})\vec{F}) = \tau\;,
\end{align}
thus restoring the expected result.

Using the closest points of the backbones for computing interactions is not without its conceptual problems:
Since the contact forces arise due to elastic deformation of the material in and around the overlap area, one would assume that increased overlap between cells should generally cause greater repulsive forces.
A simple example, illustrated in \cref{pan:model_rod_overlap}, shows that the approximation does not have this property.
A cell that is intersecting another much longer cell orthogonally with a single node will experience the same node forces even as it is pivoted around the intersecting node until the cells are nearly parallel.
This transformation does not change the closest distance between the backbones, but it can increase the overlap area by a factor close to the division length of the rotated cell. In practice, this problem results in long rod cells needing more overlap in orthogonal direction to maintain the same forces, effectively slightly reducing the hardness of the cell in the orthogonal direction.

Another related problem which is more relevant to our focus on force continuity is due to the restriction of this scheme to one interaction force between every pair of cells: Again considering a situation similar to \cref{pan:model_rod_overlap}, the closest point on the cell's backbone becomes undefined as the backbone becomes parallel to that of its interaction partner. While, physically, two cells touching along their straight sides should be a very stable configuration, the closest-point recipe always yields a definitive attack point in arbitrary positions on the backbone, inducing torques on the interacting cells which only balance over multiple time steps to keep the cells parallel. We will see below that this leads to fluctuations in the rod cell model that are absent for disk cells due to the presence of two attack points for each cell and their resulting ability to balance torques.

\subsection{Mobilities}
We use similar mobilities as for the disk model, but altered to account for the spherocylindrical shape~\cite{luders_microscopic_2021}, with
\begin{alignat}{5}
\label{eq:rod_mob_par}
\mu^\parallel_\text{rod}(a) &= \frac{1}{2\pi\,\eta\,\qty(2R\,a)}  \,\Big(&\log(a) &-0.1404 &+&\,1.034&\,a^{-1}&-&0.228 &\,a^{-2}\Big)\\
\label{eq:rod_mob_perp}
\mu^\perp_\text{rod}(a)     &= \frac{1}{4\pi\,\eta\,\qty(2R\,a)}  \,\Big(&\log(a) &+0.8369 &+&\,0.5551&\,a^{-1}&-&0.06066 &\,a^{-2}\Big)\\
\label{eq:rod_mob_rot}
\mu\st{rot}_\text{rod}(a)   &= \frac{3}{ \pi\,\eta\,\qty(2R\,a)^3}\,\Big(&\log(a) &-0.3512\,&+&\,0.7804&\,a^{-1}&-&\,0.09801&\,a^{-2}\Big)
\end{alignat}
for a cell aspect ratio $a = b_i/2R + 1$.
These mobilities are slightly higher than those in equations \eqref{eq:model_mob_par}–\eqref{eq:model_mob_rot}, and converge with them for $a = 1$.
Note that, in the case of finite internal spring hardness $H\st{int}$, which we do not consider in this study, mobilities for the internal degree of freedom are computed as a function of $\mu^\parallel$ in analogy to the disk cell model (see end of \cref{sec:force_decomposition}).

\section{Model comparison}
To test whether the disk cell model defined in \cref{sec:disks} indeed achieves the desired properties and how it compares to the alternative rod cell from \cref{sec:rods}, we carry out simulations which demonstrate the consequences of these modelling choices. In \cref{sec:forcecontinuity}, we will examine the microscopic mechanical interactions and their implications on a population level. Subsequently we turn to orientational dynamics as an example of emergent behavior.

\begin{figure}[!h]
\centering
\includegraphics{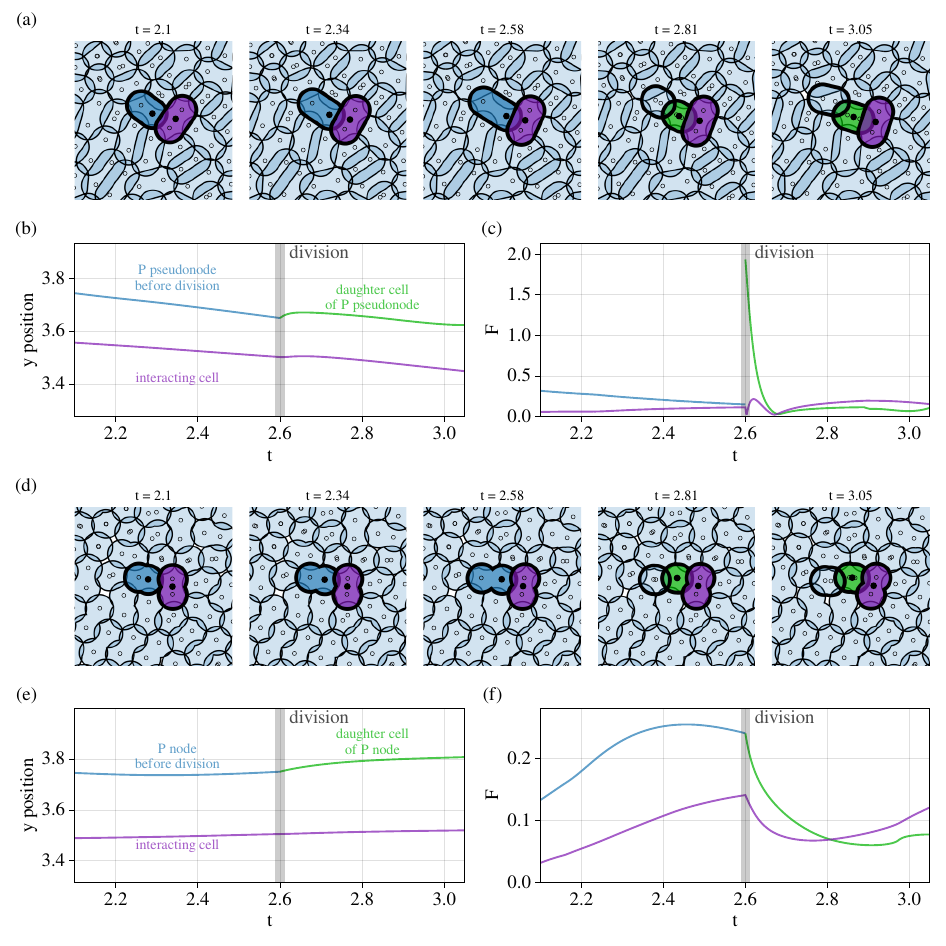}
\caption{%
Continuity of positions and forces.
\panel{pan:rodsnapshots}~Snapshots of single growing rod cell (blue) immersed in a colony of passive cells. Simulations done using a domain size of $7\cross 7$ with bi-periodic boundary conditions and parameters $R=0.5$, $\eta=0.05$, $Y=50$, $l\st{max}=2$ and $H\st{int}=\infty$ with all growth rates set to 0 except one growing particle as described in the text. Tracked objects (marked with $\newmoon$): P pseudonode of the growing cell, one cell (purple) interacting with it, and daughter cell (green) resulting from the P pseudonode. All other pseudonodes are marked with $\Circle$.
\panel{pan:rodposition}~Positions of tracked entities in the simulation in \cref{pan:rodsnapshots}.
\panel{pan:rodforces}~Forces on tracked entities in the simulation in \cref{pan:rodsnapshots}.
\panel{pan:disksnapshots}~Snapshots of single growing disk cell (blue) immersed in a colony of passive cells. Parameters and tracked objects analogous to \cref{pan:rodsnapshots}, except $H\st{int}=1$ here. Daughter cell (green) arises from the tracked P node of mother cell (blue).
\panel{pan:diskposition}~Positions of tracked entities in the simulation in \cref{pan:disksnapshots}.
\panel{pan:diskforces}~Forces on tracked entities in the simulation in \cref{pan:disksnapshots}.
}
\label{fig:forcecontinuity}
\end{figure}

\begin{figure}[!h]
\centering
\includegraphics{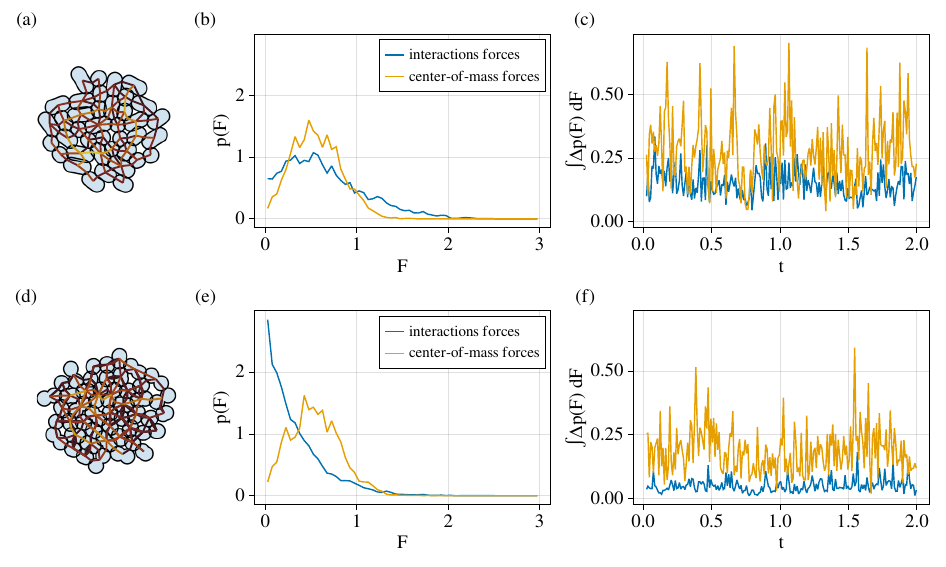}
\caption{%
Force statistics.
\panel{pan:rodclumpsnapshots}~Snapshots of a small circular colony of growing rod cells. Simulations done using a circular domain of diameter $8$ with absorbing boundary conditions and parameters $R=0.5$, $\eta=0.05$, $Y=80$, $l\st{max}=2$ and $H\st{int}=\infty$ with growth rates chosen uniformly between $0.75$ and $1.25$. Lines from cell centers to boundaries indicate intercellular interactions, where brighter colors represent stronger forces.
\panel{pan:rodforcedistrib}~Distribution of individual interaction forces and total center-of-mass forces on each cell in the simulation depicted in
\cref{onlyletter:pan:rodclumpsnapshots} averaged over 2 generations.
\panel{pan:rodforcechange}~Integrated squared difference between consecutive force distributions 0.01 generations apart.
\panel{pan:diskclumpsnapshots},
\panel{pan:diskforcedistrib},
\panel{pan:diskforcechange}~Same as in \cref{onlyletter:pan:rodclumpsnapshots,onlyletter:pan:rodforcedistrib,onlyletter:pan:rodforcechange} but for disk cells (with identical simulation parameters, except here $H\st{int}=1$).
}
\label{fig:forcestatistics}
\end{figure}

\subsection{Force continuity}
\label{sec:forcecontinuity}
To be able to track inter-particle forces on the microscopic level, we prepare a small simulation in a square domain with a side length of 7 node diameters and bi-periodic boundary conditions. We place 49 non-growing particles ($\gamma_i=0$) with random $g_i\in[0,1)$ and random $\varphi_i$ in the simulation domain, except a single selected particle $k$ near the center for which $g_k$ is set to $0.4$. After evolving the system for $2.0$ time units to relax initial mechanical stresses, the particles are in a configuration with significant overlaps, as visible in the snapshots in \cref{fig:forcecontinuity}. We then set the growth rate of the chosen particle to $\gamma_k=1$ and observe it until after its division at $t=2.6$ (blue particle with green descendant in \cref{pan:rodsnapshots,pan:disksnapshots}).

As a measure for the smoothness of the dynamics, \cref{pan:rodposition} shows the $y$ positions of the P pseudonode of the growing rod cell (blue), of the daughter cell emerging from it (green) and of one interacting cell in the vicinity (purple). While the positions have to be continuous by definition (considering the division protocol and their emergence from the integration of ODEs), a distinct kink at division is visible, indicating an abruptly changing velocity for the daughter cell. \cref{pan:rodforces} indicates a large force jump as the reason for this kink: The daughter cell is placed at the position of the mother cell's P pseudonode, resulting in an instantaneous loss of the rigid backbone's support which compensated the compressive forces from the surrounding cells. Another significant jump can be seen in the force on the interacting cell due to the instantaneous replacement of the mother cell by differently-shaped objects. Both of these forces converge back to levels typical for the overall configuration after a short time of mechanical relaxation.

In contrast, the corresponding plots for the disk cell simulation in \cref{pan:disksnapshots} show smooth node
trajectories (\cref{pan:diskposition}) and continuous forces (\cref{onlyletter:pan:diskforces}) without jumps. For the continuity between the force on the P node and that on the daughter cell, this is a result of both our choice of internal spring in \cref{eq:model_disk_internalforce}, which mimics the external repulsion after division, and the softness factor according to \cref{eq:model_disk_softness}, which avoids instantaneous quadrupling of forces when nodes are replaced by entire cells with two overlapping nodes each. The softness factor and the fact that the mother cell's nodes are replaced by objects which maintain the exact shape from before division together also guarantee force continuity for the interacting cell. Note that a mechanical relaxation process exploring the newly added degrees of freedom is also visible for disk cells, after the two compartments are untethered during division, but, in contrast to rod cells, the forces remain continuous and within the same order of magnitude (compare $y$ axis scales in \cref{onlyletter:pan:rodforces,onlyletter:pan:diskforces}).

So far, we have seen that our model definition is able to avoid large jumps for individual interaction forces when an isolated cell divides while other dynamics in the colony are limited to passive rearrangements. Next, we wanted to explore the consequences of this improvement for an entire growing colony, where fluctuations are expected in any event due to a continuous stream of rearrangements, division and removal events. This could be important for extracting physically meaningful stresses and instantaneous velocities from a simulation.

To investigate this, we set up a small simulation with circular absorbing boundary conditions, that is, cells beyond a certain distance from the center are removed, while all cells continually grow with individual random growth rates $\gamma_i$ drawn from a uniform distribution on the interval $[0.75,1.25]$. Starting with a few initial cells, the domain quickly fills and the simulations enter a dynamic steady state, which we then simulate and analyze for 2 more generations by recording all instantaneous center-of-mass forces $\vec{F}\st{cm}_i$ (as in \cref{eq:model_generic_eom_cmpos}) and all individual interaction forces between all cells every $0.01$ generations. \cref{pan:rodclumpsnapshots} and \ref{onlyletter:pan:diskclumpsnapshots} show snapshots of the steady state for disk and rod cells, respectively, with interaction forces represented as lines connecting cell centers and attack points on cell boundaries. At any instant in time, these interaction forces -- and consequently the total center-of-mass forces -- have different magnitudes, despite leading to an on-average smooth outward flow of the growing aggregate. The corresponding probability distributions of the magnitudes of these forces during the observation period of 2 generations are shown in \cref{pan:rodforcedistrib} and \ref{onlyletter:pan:diskforcedistrib}.

These distributions indicate that, in both models, cells experience a similar range and distribution of center-of-mass forces. This is consistent with the similarity of the two model definitions in terms of cell shape, size and mobility: Identical growth rates imply a similar area production rate, which in turn requires similar expansion velocities in steady state in different parts of the colony. These velocities then translate to similar average center-of-mass forces and also distributions, if the fluctuations are not larger than the spatial differences between different locations. In contrast, individual interaction forces depend on the concrete force laws employed in each model. Although they are based on Hertzian repulsion in both cases according to \cref{eq:model_disk_force,eq:model_rod_external_steric}, two rod cells can have at most one interaction force, while two disk cells can have up to four (practically: three) interactions which are, in addition, modulated by the softness factor, \cref{eq:model_disk_softness}. While the increased number should generally result of weaker overall interaction forces, \cref{pan:rodforcedistrib} also shows a difference in the shape of the interaction force distribution compared to \cref{pan:rodclumpsnapshots}, with more dominant weak force contributions and a stronger exponential-like decay towards high forces for disk cells.

The long tails in the interaction force distribution for rod cells are consistent with strong fluctuations and force peaks described before (cf. \cref{fig:forcecontinuity}). However, the former could also simply be due to the different number and attack points of the interaction forces. To test more directly whether the aberrant fluctuations also manifest on a population level or are minor compared to physical fluctuations, we calculated the integrated squared difference between consecutive force distributions 0.01 generations apart, i.e., $\Delta P^2 =\int_0^\infty\left[p_{t}(F)-p_{t-0.01}(F)\right]^2\intd F$, as shown in \cref{pan:rodforcechange} and \ref{onlyletter:pan:diskforcechange}. While the absolute magnitude of $\Delta P^2$ depends on the size of the simulation, the time interval and the resolution with which the distributions are calculated, its relative magnitude for constant numerical parameters can be taken as a measure for temporal force fluctuations. The results show that fluctuations in interaction forces are much weaker in the disk cell model compared to the rod cell model. This might not be entirely surprising given that interaction forces are distributed over a much wider range in the rod cell case and the number of interaction forces is lower, leading to stronger fluctuations. Nevertheless, we can also observe a reduction in fluctuations of the center-of-mass forces. Since these distributions were similar for both models (as well as the number of cells and, therefore, forces) and center-of-mass forces translate directly into motion, we can indeed conclude that the smoother nature of the disk cell model translates to smoother dynamics on the collective level.

\subsection{Collective dynamics, orientational order, angle statistics}
\label{sec:orientations}

\begin{figure}
\centering
\includegraphics{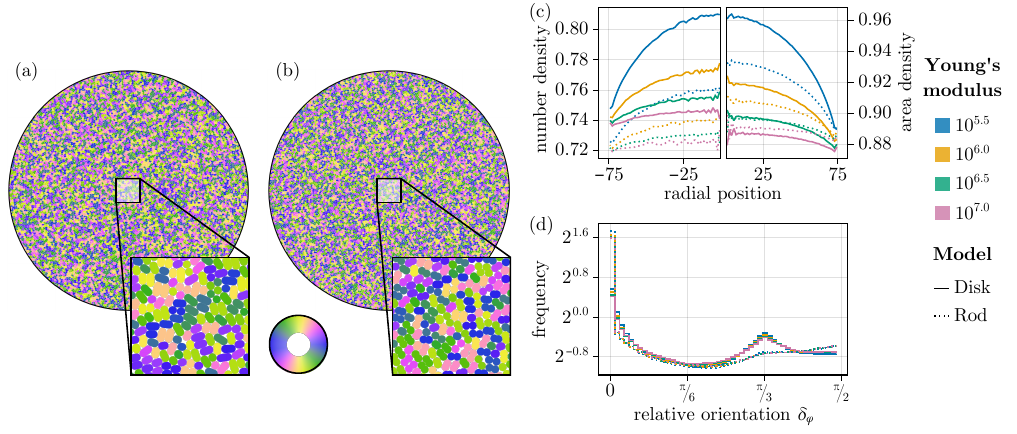}
\caption{%
\panel{pan:circles-rodsnapshot}~Snapshot of a circular colony of radius $75$, for rod cells with parameters $R = 0.5$, $Y = 10^7$, $H\st{int} = \infty$, $\eta = 0.05$ and growth rates chosen uniformly from $0.75$ to $1.25$. For easy comparison to disk cell simulations, the division length is set to $l\st{max} = 4R$. Cells are color-coded by orientation.
\panel{pan:circles-disksnapshot} Same as \cref{onlyletter:pan:circles-rodsnapshot} with disk cells. Parameters unchanged, except for $H\st{int} = 1$.
\panel{pan:circles-density}~Number density (left) and area density (right) in the steady state as a function of radial position for disk and rod model simulations (solid and dotted lines respectively) with different Young's moduli, averaged over eight realizations from $t = 15.0$ to $30.0$. All other parameters are chosen as in \cref{onlyletter:pan:circles-rodsnapshot}.
\panel{pan:circles-anglestatistics}~Frequency density of relative orientations of overlapping neighbors in the same simulations.
}
\label{fig:circles}
\end{figure}

As we keep the average growth rate constant for all particles throughout the simulation, without any form of cell removal, system size would grow exponentially forever.
Because cells close to the center of the system need to work against the friction of all cells downstream in order to expand, stresses in the center of the colony increase with system size~\cite{isensee_stress_2022}.
This increase in central stress cannot be compensated for by the Hertzian interaction forces, which have a finite upper limit at full overlap $2R$.
Therefore, the central density would increase steadily until at some point the steric interactions would be overwhelmed by stress, causing the center of the colony to collapse.
To avoid this increase in density, we  keep the system size limited by removing all cells with a center of mass position more than 75 node diameters from the system center, thus forming circular colonies of constant radius.
This allows the colony to form a steady state of constant radial expansion flow with growth compensated by the removal of cells at the boundary.

We initialize our simulations with four cells of random orientation and growth progress placed in the center of the system and letting the colony expand until it densely fills the entire domain (at about 12 generations).
We then simulate this steady state until reaching $t = 30~\text{generations}$.
For the analysis, we discard the first 15 generations to ensure that the transient expansion phase is excluded.
For better statistics, we run these simulations eight times for each parameter set, with different random seeds and initial conditions for every realization.
\cref{pan:circles-rodsnapshot} and \ref{onlyletter:pan:circles-disksnapshot} show the final state of a simulation for rod and disk cells respectively.

The main difference between the disk and rod model are their steric interaction laws.
Therefore, we first compare the steady-state densities at equal Young's modulus $Y$.
\cref{pan:circles-density} shows the time-averaged number density profile for disks and rods as a function of radial position, measured by counting cell centers of mass within concentric bins of width $1.5$.
We observe a parabolic profile, taking its maximum at the domain center and falling off towards the outer edge.
As expected, peak density decreases with increasing Young's modulus for both cells, while density at the absorbing boundary, where stresses are low, changes only little.
Number densities for the rod model are consistently lower than those for disk cells of equivalent radius.

To see how much of this is caused by simply the difference in area between disk and rod cells of equal growth progress, we additionally compute area densities by adding the individual cell areas in a bin and dividing by the bin area.
We find that while for lower Young's moduli the area density is still lower in the rod simulations, the difference between models decreases with increasing hardness.
For $Y=10^7$, area density for the disk model is in fact slightly lower than for the rod model everywhere in the domain, while number density is still significantly higher.
This implies that while the area difference does contribute to the observed number density difference, for softer cells there must be additional effects, such as higher overlaps or more efficient packing, that affect both density measures.

To assess the influence of the different cell shapes on the microscopic structure of a dense colony, we measure the relative orientation $\delta_\varphi$ between overlapping cells in these simulation and compute a frequency histogram, which is shown in \ref{pan:circles-anglestatistics}.
For both models, we see a distinct peak for parallel cells ($\delta_\varphi = 0$), corresponding to local nematic alignment.
This is peak is much stronger for the rod model, with almost twice as many parallel pairs as for disk cells.
Intuitively, this can be explained by the flat sides of the rods, which enable and stabilize parallel cell configuration.

For both models, orientations between $0$ and around $\pi/3$ occur with the lowest frequency.
At larger relative angles, the frequency increases again, with a pronounced secondary peak at $\pi/3$ for the disk model.
For even higher $\delta_\varphi$, the histogram plateaus for disk cells, while rising further towards $\pi/2$ for the rod model.
The secondary peak at $\pi/3$ observed in the disk data could be interpreted as the signature of local quasi-hexagonal arrangements, in which nodes of one cell can interact with the indentation between the nodes of another.
In any case, these differences are manifestations of different steric interaction rules for the two models.

\begin{figure}
\centering
\includegraphics{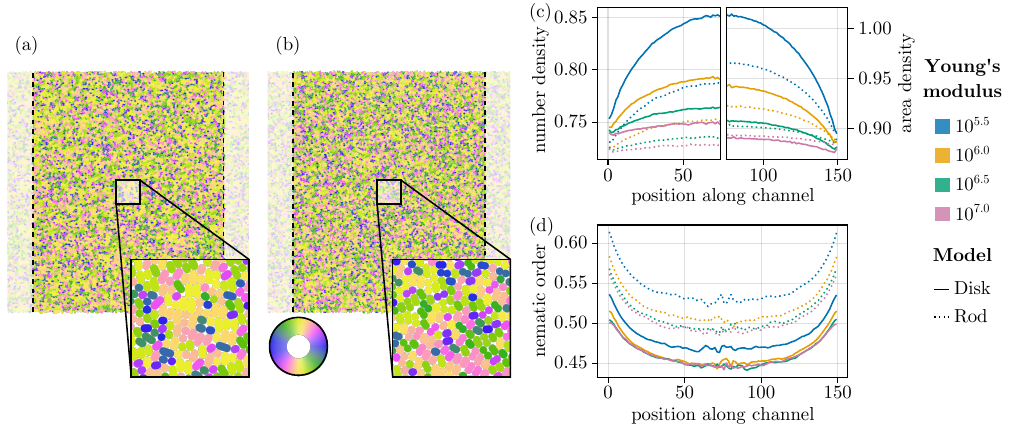}
\caption{%
\panel{pan:channels-rodsnapshot} Snapshot of an x-periodic channel of height $150$ and width $120$ at $t = 30$ generations, for rod cells with parameters $R = 0.5$, $Y = 10^7$, $H\st{int} = \infty$, $\eta = 0.05$ and growth rates chosen uniformly from $0.75$ to $1.25$. For easy comparison to disk cell simulations, the division length is set to $l\st{max} = 4R$. Cells are color-coded by orientation.
\panel{pan:channels-disksnapshot} Same as \cref{onlyletter:pan:channels-rodsnapshot} with disk cells. Parameters unchanged, except for $H\st{int} = 1$.
\panel{pan:channels-density}~Number density (left) and area density (right) in the steady state as a function of position along the channel axis for disk and rod model simulations (solid and dotted lines respectively) with different Young's moduli, averaged over eight realizations from $t = 15.0$ to $30.0$. All other parameters are chosen as in \cref{onlyletter:pan:channels-rodsnapshot}.
\panel{pan:channels-orderparameter} Nematic order parameter $\xi$ as a function of position along the channel axis for the same simulations
}    \label{fig:channels}
\end{figure}

Aside from these local alignment effects, it is known from literature that dense growing systems also feature nematic order at larger scales.
For particles with more extreme aspect ratios, studies report the emergence of large locally aligned patches of microdomains~\cite{you_geometry_2018, langeslay_microdomains_2023}.
Due to the fairly low anisotropy of our particles, we cannot observe these larger structures here (\cref{pan:circles-rodsnapshot} and \ref{onlyletter:pan:circles-disksnapshot}).

Another effect observed in growing nematic particles is a tendency to align with shear flows.
Since the expansion flow in circular colonies is shear free if the colony is sufficiently close to incompressibility~\cite{dellarciprete_growing_2018}, shear-induced alignment is not visible in these systems.
We therefore alter our simulation setup to induce a strong uniaxial shear flow throughout the entire colony, by replacing the radially symmetric absorbing boundaries with axis-parallel absorbing boundaries in one direction and confining boundaries in the other.
The global nematic alignment resulting from this rectangular channel geometry has been observed and studied from different perspectives before\cs\cite{volfson_biomechanical_2008, orozco-fuentes_order_2013, you_confinementinduced_2021, isensee_stress_2022}.

Here, we simulate channels of size $120\times150$ units, with periodic boundaries in the x-direction and absorbing boundaries in the y-direction (\cref{pan:channels-rodsnapshot} and \ref{onlyletter:pan:channels-disksnapshot}).
We use the same protocol as in the previous simulations and again discard the transient phase of the colony evolution.
We repeat our previous measurements of number and area density (\cref{pan:channels-density}) and find that results are qualitatively similar, although densities are higher than in the circular systems.
This can be understood as a consequence of the confinement: stresses can now only be relaxed in the unconfined direction.

For both models, we can observe the expected nematic alignment with the channel direction throughout the system.
This is directly visible in \cref{pan:channels-rodsnapshot} and \ref{onlyletter:pan:channels-disksnapshot} where cells are color-coded by orientation.
We quantify this with a nematic order parameter $\xi = \abs{\langle\exp(2i \varphi_j)\rangle_j}$, where the average is over individual cells.
This order parameter is $0$ for isotropically distributed $\varphi_i$, and $1$ for identical angles.
We measure $\xi$ along the channel axis, averaging over bins of height $1.5$ that span the width of the channel.
\cref{pan:channels-orderparameter} shows that the order parameter is around $0.45..0.6$ for both models, corresponding to the visual impression of a distinct alignment bias with remaining disorder.
For both models $\xi$ is higher near the outlets and falls to a plateau in the center of the colony.
$\xi$ is systematically higher for the rod model than the disk model, which is consistent with the increased preference for local nematic alignment observed in the circle domain.

With increasing Young's modulus, the order parameter profiles are shifted to lower values, while the difference between outlet and center remains similar.
Interestingly, we observe that while the density profiles still change significantly even between the two highest Young's moduli we simulate ($Y = 10^{6.5}$ and $10^7$), the order parameter profiles change very little and appear to converge.

\section{Discussion}
In this study, we introduced a minimal model for growing and dividing cells with a focus on mechanical consistency and continuity. While the smallest mechanical unit of our model has radial symmetry (like in many other models in the literature), we explicitly combined two of these units as nodes of an entire \emph{disk cell} in order to achieve force continuity across divisions, internally between the two separating compartments as well as with interaction partners. While it is clear a priori that this approach leads to smoother behavior compared to models with instantaneous replacement of the mother cell by differently shaped children, we showed that it is also able to eliminate unphysical force jumps upon division in comparison to an established model  with incremental elongation and similar cell shape. The elimination of these jumps also led to reduced fluctuations in an entire cell colony, while larger simulations revealed similar collective dynamics. This should make it possible in the future to focus on physically necessary forces and fluctuations, for example in stress calculations\cs\cite{das_local_2019,isensee_stress_2022}, or when cell activity itself is regulated by its mechanical environment, for example pressure\cs\cite{Basan2009,pollack_competitive_2022}.

In particular, this is true for the analysis of center-of-mass trajectories of individual cells on short time scales. First of all, large unphysical displacements compensating for model discontinuities would disrupt displacement statistics such as mean-squared displacements, van Hove functions or velocity correlations. In addition, however, extracting statistically fair trajectories from arbitrary observation windows is only possible if cell identities are not lost and compartments can be tracked across divisions, which is possible in the disk cell model presented here due to the smooth transition from intracellular nodes to entire cells.

In this context, it is useful to realize that the disk cell model setup avoids another potential pitfall when analyzing the collective dynamics from an active-matter perspective: If one is interested in mechanics, i.e. the interplay of different stress/force contributions in the system, one has to account for \emph{all} forces, including active contributions. In motile active matter, a prescribed active velocity $v_a$ can often simply be converted to an equivalent active force $F_a=v_a/\mu$ via the translational mobility $\mu$, but the same is conceptually difficult with proliferating systems, not to mention models with instantaneous replacement where no actual motion can be associated with the division process. Even in the models with continuous expansion discussed here, this difficulty left its footprint in that certain choices had to be made regarding the internal mobility for disk cells (see \cref{sec:force_decomposition}) and how to account for the forces associated with a prescribed backbone length for rod cells (see \cref{sec:rods_growth_and_division}). However, at least in the new disk cell model, once these choices have been made, all motion is caused by explicitly calculated forces, including the active elongation force of the internal spring, eliminating the need to account for it separately.

It should be emphasized that we view our disk cell model more as a general-purpose model for multicellular systems rather than for a particular cell type. Radially symmetric interaction potentials for individual nodes were chosen for simplicity, with the transient dumbbell-shape during elongation as a necessary consequence, and additional modifications were only introduced in order to achieve mechanical consistency. However, there are biological examples, such as coccus-shaped bacteria, where growth and division closely follow the dynamics introduced here for our disk cell model\cs\cite{WelkerMaier2018}, as well as counterexamples such as snapping division, where discontinuities might be physical at a certain level of description\cs\cite{Zhou2016}. As we have seen in \cref{sec:orientations}, the collective dynamics in terms of orientational order and packing closely resemble those of short rod cells, independent of the exact shape. Given the unavoidable anisotropy involved in even the simplest symmetric division process from 1 to 2 cells, our model therefore represents a general smooth alternative to generic cells which exhibit anisotropy upon division.

Here, we chose to tie the equilibrium backbone length in \cref{eq:model_disk_restlength} directly to growth progress, such that the elongation of a cell happens in the most simple linear fashion throughout the lifetime of the cell. However, for more detailed studies, the two nodes of our model cell could be seen as analogs of the two separating nuclei during mitosis or the two separating compartments during cytokinesis, since they both potentially involve the exertion of forces on the environment through various active cytoskeleton components. Other choices for periods of length/shape change through the cell's life cycle are therefore conceivable in analogy to cell cycle phases and could be easily implemented.

Besides these alternative elongation dynamics and the regulated growth considered above, there are a number of other features which would be straight forward to implement as extensions but weren't considered in this study. This includes thermal noise (positional, orientational or in growth-progress space) in analogy to active Brownian particles, asymmetric division and the extension to 3 dimensions. By serving as the ``mechanical backbone'' for extensions with additional mechanical, chemical or regulatory features, we hope that our model will facilitate the characterization of growing multicellular systems in the framework of active matter and non-equilibrium statistical physics.

\section*{Acknowledgements}
The authors would like to thank W. Till Kranz for early ideas on incorporating division into particle-based models and for helpful comments on the manuscript. We acknowledge support from the Max Planck School Matter to Life and the MaxSynBio Consortium which are jointly funded by the Federal Ministry of Education and Research (BMBF) of Germany and the Max Planck Society.

\section*{Software Availability}
Reference implementations in the Julia programming language\cs\cite{bezanson_julia_2017} of both models presented here (with several additional features) are available via the package \href{http://biome.inparts.org}{InPartSBiome.jl} at \url{http://biome.inparts.org} or through the persistent identifier in Ref.~\citenum{InPartSBiome}, together with example simulation scripts. These implementations are based on \href{http://inparts.org}{InPartS.jl}, available at \url{http://inparts.org} or through the persistent identifier in Ref.~\citenum{InPartS}, which is developed in our group and provides a common modeling framework for \textbf{In}teracting \textbf{Part}icle \textbf{S}imulations with ready-made tools for the implementation of various mechanical interaction laws, out-of-the-box import/export and visualization.

\end{document}